\documentclass[12pt,eqsecnum,amsfonts,aps]{revtex4}

\usepackage{graphicx}

\usepackage{bm}

\newcommand{\beq}{\begin{equation}}
\newcommand{\eeq}{\end{equation}}
\newcommand{\beqs}{\begin{eqnarray}}
\newcommand{\eeqs}{\end{eqnarray}}

\begin{document}

\baselineskip 6.0mm 

\title{Some Exact Results on Bond Percolation} 

\bigskip

\author{Shu-Chiuan Chang$^a$ and Robert Shrock$^b$}

\affiliation{(a) \ Physics Department \\ 
National Chen Kung University,  \\
Tainan 70101, Taiwan}

\affiliation{(b) \ C. N. Yang Institute for Theoretical Physics \\
Stony Brook University \\
Stony Brook, NY 11794 }

\bigskip
\bigskip

\begin{abstract}

We present some exact results on bond percolation.  We derive a relation that
specifies the consequences for bond percolation quantities of replacing each
bond of a lattice $\Lambda$ by $\ell$ bonds connecting the same adjacent
vertices, thereby yielding the lattice $\Lambda_\ell$. This relation is used to
calculate the bond percolation threshold on $\Lambda_\ell$. We show that this
bond inflation leaves the universality class of the percolation transition
invariant on a lattice of dimensionality $d \ge 2$ but changes it on a
one-dimensional lattice and quasi-one-dimensional infinite-length strips.  We
also present analytic expressions for the average cluster number per vertex and
correlation length for the bond percolation problem on the $N \to \infty$
limits of several families of $N$-vertex graphs.  Finally, we explore the
effect of bond vacancies on families of graphs with the property of bounded
diameter as $N \to \infty$.

\end{abstract}


\maketitle


\pagestyle{plain}
\pagenumbering{arabic}

\newpage

\section{Introduction}
\label{intro}

Let $G=G(V,E)$ be a connected graph defined by a set $V$ of vertices ( = sites)
and a set $E$ of edges (= bonds) connecting pairs of vertices.  Bond
percolation is an interesting geometrical phenomenon in which one considers $G$
to be modified in such a manner that each bond is independently randomly
present with probability $p$. We denote the number of vertices and bonds as
$N=N(G)=|V|$ and $e(G)=|E|$, and the formal limit of $G$ as $N \to \infty$ as
$\{G\}$ \cite{loops}. In the context of statistical mechanics, one is often
interested in the $N \to \infty$ limit of a regular lattice graph $G$, denoted
$\Lambda$. As the bond occupation probability $p$ decreases from 1, there are
more and more absent bonds on $\Lambda$, and the probability $P(p)_\Lambda$
that a given vertex belongs to an infinite cluster decreases monotonically
until, at a critical value, $p_{c,\Lambda}$, it vanishes and remains
identically zero for $0 \le p < p_{c,\Lambda}$. This value, $p_{c,\Lambda}$, is
the critical bond occupation probability, also called the critical bond
percolation threshold. Other quantities also behave nonanalytically at
$p_{c,\Lambda}$.  For example, the average cluster size, $S(p)_\Lambda$,
increases monotonically as $p$ increases from 0 and diverges as $p$ approaches
$p_{c,\Lambda}$ from below.  Thus, the percolation transition is a geometrical
transition from a region $0 \le p < p_{c,\Lambda}$ in which only finite
connected clusters exist, to a region $p_{c,\Lambda} \le p \le 1$ in which
there is a percolating cluster containing an infinite number of vertices and
bonds.  Another interesting quantity is the average number of (connected)
clusters, including single vertices, divided by the total number of lattice
vertices, in the limit $N \to \infty$, denoted as $\langle n \rangle_\Lambda$.
Analogous statements hold for site percolation, in which each vertex of
$\Lambda$ is independently randomly present with probability $p$. Methods for
studing percolation have included exact mappings, series expansions, Monte
Carlo simulations, and the renormalization group.  In addition to its intrinsic
interest, percolation gives insight into a number of important phenomena such
as the passage of fluids through porous media, electrical currents through
composite materials consisting of conducting and insulating components, and the
effect of lattice defects and disorder on thermal critical phenomena. Some
reviews include \cite{essam80}-\cite{math}.

A basic aspect of the percolation transition on a lattice $\Lambda$ or, more
generally, a limit $\{ G \}$ of a family of graphs, concerns the dependence of
the critical behavior and the value of $p_{c, \{ G \}}$ on properties of $\{ G
\}$.  It is known that the percolation transition is in the universality class
of the $q=1$ ferromagnetic Potts model, and that its upper critical
dimensionality is $d_u=6$.  For a given lattice dimensionality $d$ above the
lower critical dimensionality, $d_\ell=1$, the universality class of the
percolation transition is independent of details of the lattice structure, such
as the coordination number. In contrast, other quantities, such as the critical
bond percolation value $p_{c,\Lambda}$, are non-universal and do depend on
lattice properties like coordination number \cite{stauffer}-\cite{parv07}.  The
percolation threshold $p_{c,\Lambda}$ is usually a decreasing function of the
lattice coordination number, although certain exceptions have been found
\cite{vandermarck,wierman02}.

In this paper we report some exact analytic results concerning the factors that
determine the critical behavior of the percolation transition and the
dependence of percolation quantities on the properties of $\{ G \}$.
Specifically, we derive a relation that specifies the consequences for bond
percolation quantities of replacing each bond of a lattice $\Lambda$ by $\ell$
bonds connecting the same adjacent vertices.  This result elucidates the effect
of an arbitrarily great increase in the degrees of the vertices of $G$ and its
$N \to \infty$ limit, $\{ G \}$. (Here, the degree of a vertex is defined as
the number of bonds connecting to this vertex.)  We show that this $\ell$-fold
bond inflation leaves the universality class of the percolation transition
invariant on a lattice of dimensionality $d \ge 2$ but changes it on a
one-dimensional lattice and on quasi-one-dimensional infinite-length,
finite-width lattice strips.  This is demonstrated by changes in the critical
exponents governing the divergences in the correlation length and average
cluster size as $p \nearrow 1$.  We also present analytic expressions for the
average cluster number per vertex, $\langle n \rangle$, for some families of
graphs containing repeated complete graphs $K_r$ \cite{kr} as subgraphs.  As
part of this, we discuss the analytic properties of $\langle n \rangle$ in the
complex $p$ plane and use these to determine the radius of convergence of the
relevant small-$p$ series expansion of $\langle n \rangle$.  Finally, we
explore the effects of bond vacancies on families of graphs with the property
of bounded diameter \cite{diam} as $N \to \infty$.  The present work extends
our previous study in \cite{pc}.


\section{General Background and Calculational Methods} 

In this section we discuss the calculational methods that we employ.  We make
use of the fact that several quantities in bond percolation can be
obtained from the partition function for the $q$-state Potts model
\cite{essam80,fk,wu78,wurev} in the limit $q \to 1$. We review this connection
next.  In thermal equilibrium at temperature $T$, the general Potts model
partition function in an external magnetic field $H$ is given by
$Z = \sum_{ \{ \sigma_n \} } e^{-\beta {\cal H}}$ with the Hamiltonian
\beq
{\cal H} = -J \sum_{\langle i j \rangle} \delta_{\sigma_i, \sigma_j}
- H \sum_\ell \delta_{\sigma_\ell,1} \ ,
\label{ham}
\eeq
where $i, \ j, \ \ell$ label vertices of $G$, $\sigma_i$ are
classical spin variables on these vertices taking values in the set
$I_q = \{1,...,q\}$, $\beta = (k_BT)^{-1}$, and $\langle
i j \rangle$ denote pairs of adjacent vertices. 

The zero-field Potts model Hamiltonian ${\cal H}$ and partition function $Z$
are invariant under the global transformation in which $\sigma_i \to g \sigma_i
\ \forall \ i \in V$, with $g \in S_q$, where $S_q$ is the symmetric (=
permutation) group on $q$ objects.  Because of this invariance, one can,
without loss of generality, choose the value picked out by the external 
magnetic field $H$ to be $\sigma_\ell=1$, as we have done in (\ref{ham}). 
It will be convenient to introduce the notation
\beq
K = \beta J \ , \quad h = \beta H \ , \quad a = e^K, \quad v = a-1 \ , 
\quad w=e^h \ .
\label{kdef}
\eeq

Given a graph $G=(V,E)$, a spanning subgraph $G' \subseteq G$ is defined as the
subgraph containing the same set of vertices $V$ and a subset of the bonds of
$G$; $G' = (V,E')$ with $E' \subseteq E$. We denote the number of connected
components of $G$ and $G'$ as $n(G)$ and $n(G')$, respectively.  The property
that $G$ is connected is the statement that $n(G)=1$.  The link between
percolation and the Potts model stems from the property that the Potts model
partition function can be expressed as a purely graph-theoretic sum over
contributions from spanning subgraphs $G'$. For $H=0$, this expression is 
\cite{fk}
\beq
Z(G,q,v) = \sum_{G' \subseteq G} v^{e(G')} \ q^{n(G')} \ .
\label{fk}
\eeq
As is evident from (\ref{fk}), $Z(G,q,v)$ is a polynomial in $q$ and $v$.  This
expression also allows one to generalize $q$ from positive integers to real
numbers.

As an example of the calculation of percolation quantities from (\ref{fk}), we
consider the average number of clusters per vertex.  On a graph $G$, this is
given by %
\beqs
\langle n \rangle & = & \frac{(1/N)\sum_{G^\prime} n(G^\prime)
p^{e(G^\prime)}(1-p)^{e(G)-e(G^\prime)} }{
\sum_{G^\prime} p^{e(G^\prime)}(1-p)^{e(G)-e(G^\prime)} } \cr\cr
& & \cr\cr
& = &
\frac{(1/N)\sum_{G^\prime} n(G^\prime)[p/(1-p)]^{e(G^\prime)}}{
\sum_{G^\prime} [p/(1-p)]^{e(G^\prime)}} \ .
\label{k}
\eeqs
This follows because each $G^\prime$ contains $n(G^\prime)$ connected
components, and appears in the numerator of the expression in the first line
with weight given by $p^{e(G^\prime)}(1-p)^{e(G)-e(G^\prime)}$, since the
probability that all of the bonds in $G^\prime$ are present is
$p^{e(G^\prime)}$ and the probability that all of the other $e(G)-e(G^\prime)$
bonds in $G$ are absent is $(1-p)^{e(G)-e(G^\prime)}$.  This sum in the
numerator over the set of spanning subgraphs $G^\prime$ is normalized by the
indicated denominator and by an overall factor of $1/N$ to obtain the average
number of connected components (clusters) per vertex.

On a finite graph $G$ one defines the (reduced) free energy per vertex of the
Potts model as
\beq
f(G,q,v) = \ln [ Z(G,q,v)^{1/N} ]
\label{fn}
\eeq
and, in the limit $N \to \infty$,
\beq
f(\{G\},q,v) = \lim_{N \to \infty} f(G,q,v) \ , 
\label{f}
\eeq
If one sets 
\beq
v = v_p \equiv \frac{p}{1-p} \ , \quad i.e., \ \ p = \frac{v}{v+1} = 1-e^{-K} \
, 
\label{vp}
\eeq
differentiates $f(G,q,v_p)$ with respect to $q$, and then sets $q=1$, one 
obtains $\langle n \rangle_n$, as given in eq. (\ref{k}), i.e.,
\beq
\langle n \rangle_N = \frac{\partial f(G,q,v_p)}{\partial q} \bigg |_{q=1} \ .
\label{kdfdqn}
\eeq
In particular, in the limit $N \to \infty$,
\beq
\langle n \rangle = \frac{\partial f(\{G\},q,v_p)}{\partial q} \bigg |_{q=1} 
\ .
\label{kdfdq}
\eeq

The formula (\ref{kdfdq}) relates a geometric property of (the $N \to \infty$
limit of) a bond-diluted graph with the derivative of the reduced free energy
of the zero-field Potts model, evaluated at a certain temperature, as $q \to
1$, on a graph with no bond dilution.  The mapping (\ref{vp}), in conjunction
with eq. (\ref{kdfdq}), formally associates the interval $0 \le p \le 1$ with
the interval $0 \le v \le \infty$, which is the physical range of values of the
temperature variable $v$ for the ferromagnetic Potts model, with $p \to 0^+$
corresponding to temperature $T \to \infty$, i.e., $v \to 0^+$, and 
$p \to 1$ to $T \to 0$, i.e, $v \to \infty$.  

For the case of nonzero field, denote the connected subgraphs of a spanning
subgraph $G'$ as $G'_i$, $i=1,..,n(G')$.  Then one can obtain a generalized
expression for the partition function as a sum over contributions from spanning
subgraphs, as \cite{wu78}
\beq
Z(G,q,v,w) = \sum_{G' \subseteq G} v^{e(G')} \
\prod_{i=1}^{n(G')} \Big ( q-1 + w^{N(G'_i)} \Big ) \ .
\label{clusterws1}
\eeq
Some general properties of $Z(G,q,v,w)$ were derived and exact results for
families of graphs were given in \cite{hl}-\cite{phs2}.  For the zero-field
special case, this reduces to $Z(G,q,v,1) \equiv Z(G,q,v)$. 

Extending the zero-field definition, we define the dimensionless reduced free
energy of the Potts model in an external field as
\beq
f(\{G\},q,v,w) = \lim_{N \to \infty} \frac{1}{N} \ln[Z(G,q,v,w)] \ . 
\label{fw}
\eeq

For a given $\{ G \}$, the quantities $P(p)$ and $S(p)$ can then be determined
in terms of derivatives of this reduced free energy:
\beqs
P(p) & = & 1 + \frac{\partial^2 f}{\partial h \partial q}
\Big |_{q=1, \ h = 0^+} = 1 + w\frac{\partial^2 f}{\partial w \partial q}
\Big |_{q=1, \ w = 1^+} \cr\cr
& &
\label{P}
\eeqs
and
\beqs
S(p) & = & \frac{\partial^3 f}{(\partial h)^2 \partial q}
\Big |_{q=1, \ h = 0^+} \cr\cr
& & \cr\cr
     & = & w \frac{\partial}{\partial w} \bigg [
w \frac{\partial}{\partial w} \bigg (\frac{\partial f}{\partial q}\bigg )
\bigg ] \bigg |_{q=1, \ w = 1^+} \ . 
\label{S}
\eeqs
where $h=0^+$ and $w=1^+$ mean $\lim_{h \to 0^+}$ and $\lim_{w \to 1^+}$. 

We recall the forms of the singularities in some quantities at the percolation
transition. On a given lattice $\Lambda$, as the
bond occupation probability increases from 0 to 1, $\langle n \rangle_\Lambda$
decreases monotonically from 1 to 0 and has a singularity at the critical
threshold value $p=p_{c,\Lambda}$ of the form 
\beq
\langle n \rangle_{\Lambda, sing.} \sim |p-p_{c,\Lambda}|^{2-\alpha} \quad 
{\rm as} \ \ p \to p_{c,\Lambda} \ . 
\label{nsing}
\eeq
As $p$ decreases from 1 to $p_{c,\Lambda}$, the average probability that a
vertex is connected to the infinite cluster, $P(p)_\Lambda$, decreases to zero
like
\beq
P(p)_\Lambda \propto (p-p_{c,\Lambda})^\beta \quad {\rm as} \ \ p \searrow
p_{c,\Lambda} \ . 
\label{betadef}
\eeq
As $p$ increases from 0 toward $p_{c,\Lambda}$, the average size of a
percolation cluster, $S(p)_\Lambda$, diverges like
$1/(p_{c,\Lambda}-p)^{\gamma'}$.  As $p$ decreases toward $p_{c,\Lambda}$ from
above, $S(p)_\Lambda$, defined as the average size of finite clusters, diverges
like $1/(p-p_{c,\Lambda})^{\gamma}$. Scaling and renormalization group methods
yield $\gamma=\gamma'$, so that
\beq
S(p)_\Lambda \propto \frac{1}{|p-p_{c,\Lambda}|^\gamma} \quad {\rm as} \ \ p 
\to p_{c,\Lambda} \ . 
\label{spsing}
\eeq
Similarly, one can define a correlation length $\xi(p)$ characterizing the size
of clusters, and this diverges at the percolation transition like 
\beq
\xi(p) \propto \frac{1}{|p-p_{c,\Lambda}|^\nu} \quad {\rm as} \ \ p 
\to p_{c,\Lambda} \ . 
\label{xidiv}
\eeq
In the Potts model, the divergence in the correlation length corresponds to the
equality of the leading and subleading eigenvalues of the transfer matrix, 
\beq
\xi = \frac{1}{\ln(\lambda_{max}/\lambda_{submax})} \ . 
\label{xi}
\eeq
Setting $q=1$ and $v=v_p$ in Eq. (\ref{vp}), one thus determines the divergence
in $\xi(p)$ for the bond percolation problem as $p \to p_{c,\Lambda}$ from the
divergence in $\xi$ for the ferromagnetic Potts model.

The relation of percolation to the $q=1$ Potts ferromagnet implies, in
particular, that the percolation transition on two-dimensional lattices is in
the same universality class as the two-dimensional $q=1$ Potts ferromagnet,
with the exactly known exponents $\alpha=-2/3$, $\beta=5/36$, $\gamma=43/18$,
$\nu=4/3$, etc. \cite{wurev}.  In the context of conformal algebra, these
critical exponents are associated with a rational conformal field theory with
central charge $c=0$ (e.g., \cite{cft}).  The percolation transition on a
$d$-dimensional lattice with $d \ne 2$ is in a different universality class. In
particular, as noted before, the upper critical dimensionality for the
percolation transition is $d_u=6$, corresponding to a Ginzburg-Landau function
with highest power $\phi^3$ and critical exponents $\alpha=-1$, $\beta=1$,
$\gamma=1$, and $\nu=1/2$.  This relation is embodied in the exact solution on
the Cayley tree (reviewed, e.g., in \cite{essam80}).

We denote the critical point in $K$ of the $q$-state Potts ferromagnet on the
lattice $\Lambda$ as $K_{c,q,\Lambda}$, and the corresponding value of $v$ as
$v_{c,q,\Lambda}$. We also denote $K_{c,1,\Lambda} \equiv \lim_{q \to 1}
K_{c,q,\Lambda}$.  The critical percolation occupation probability, i.e., the
percolation threshold, $p_{c,\Lambda}$, is determined in terms of the critical
point of the corresponding $q \to 1$ Potts model as $p_{c,\Lambda} = 1 -
e^{-K_{c,1,\Lambda}}$. Exact values of $p_{c,\Lambda}$ for bond percolation on
the square (sq), triangular (tri), and honeycomb (hc) lattices were obtained in
\cite{se64} (reviews include \cite{essam80}-\cite{math}).  For later reference,
these well-known critical bond percolation thresholds are $p_{c,sq} = 1/2$,
$p_{c,tri} = 2\sin(\pi/18) = 0.347296..$, and $p_{c,hc} = 1 - p_{c,tri} = 1 -
2\sin(\pi/18) = 0.652704..$ Much work has been done on determinations of
$p_{c,\Lambda}$ values for various lattices \cite{essam80}-\cite{parv07}.  For
example, for the simple cubic (sc) lattice, $p_{c,sc} = 0.2488126(5)$
\cite{ziff98}, where the number in parentheses is the uncertainty in the last
digit.  After reviewing this background, we now proceed to our new results.


\section{Bond Percolation Quantities on Bond-Inflated Lattices}

\subsection{Basic Result} 

Consider an arbitrary graph $G = (V,E)$. We define the graph $G_\ell$ as the
graph obtained by replacing each bond of $G$ by $\ell$ bonds connecting the
same vertices.  Similarly, in the $N \to \infty$ limit, we define $\{ G
\}_\ell$ in the same manner.  A $\kappa$-regular graph $G$ is defined as a
graph all of whose vertices have the same degree (coordination number),
$\kappa$.  To cover the case of graphs that are not $\kappa$-regular, it will
also be useful to define an effective vertex degree in the $N \to \infty$
limit, namely 
\beq
\kappa_{eff}(\{G \}) = \lim_{N \to \infty} \frac{2e(G_m)}{N} \ , 
\label{kappaeff}
\eeq
For example, this can be defined on duals of Archimedean lattices \cite{wn}. 
Clearly, for a $\kappa$-regular graph $G$, $\kappa_{eff} = \kappa$. For a
$\kappa$-regular graph $G$, the $\ell$-fold bond inflation increases the vertex
degree to $\ell \, \kappa$, and similarly, for graphs that have vertices of
different degrees, the degree of each vertex is increased by the factor $\ell$.
These cases are subsumed in the $N \to \infty$ limit, as 
\beq
\kappa_{eff}(\{G \})_\ell = \ell \, \kappa_{eff}(\{G \}) \ .
\label{kapparel}
\eeq
In particular, if $\{ G \}$ is a regular lattice $\Lambda$, we denote the
lattice $\Lambda_{\ell}$ as the result of replacing each bond on $\Lambda$ by
$\ell$ bonds connecting the same adjacent vertices.

In this section we derive exact relations that express quantities
characterizing bond percolation on $\{ G \}_\ell$ in terms of the corresponding
quantities on the original $\{ G \}$ with a transformed value of the bond
occupation probability.  The starting point of our derivation is the
observation that the effect of $\ell$-fold bond inflation is embodied in the
relation
\beq
Z(G_\ell,q,v) = Z(G,q,v_\ell) \ , 
\label{zellrel}
\eeq
where
\beq
v_\ell = (v+1)^\ell-1 \ . 
\label{vell}
\eeq
Some other implications of this are discussed in \cite{ha}.  Because the
interaction of the external magnetic field with the spins $\sigma_i$ in
(\ref{ham}) is unaffected by the bond inflation, the generalization of
Eq. (\ref{zellrel}) to the case $H \ne 0$ is immediate:
\beq
Z(G_\ell,q,v,w) = Z(G,q,v_\ell,w) \ . 
\label{zellhrel}
\eeq
For our particular application, we set $v=v_p$ in Eq. (\ref{vp}), which yields
the relation
\beq
p_\ell = 1 - (1-p)^\ell = p \, \sum_{j=1}^\ell {\ell \choose j}(-p)^{j-1} 
\ , 
\label{pell}
\eeq
where ${\ell \choose j} \equiv \ell!/[j!(\ell-j)!]$ is the binomial
coefficient. The first few explicit cases, aside from $p_1 = p$, are
\beq
p_2 = p(2-p) \ , 
\label{p2}
\eeq
\beq
p_3 = p(3-3p+p^2) \ , 
\label{p3}
\eeq
\beqs
p_4 & = & p(4-6p+4p^2-p^3) \cr\cr
    & = & p(2-p)(2-2p+p^2) \ , 
\label{p4}
\eeqs
and so forth for higher $\ell$. 

From Eq. (\ref{pell}) it follows that the $r$'th derivative of $p_\ell$ with
respect to $p$ is 
\beq
\frac{d^r p_\ell}{dp^r} = (-1)^{r-1} \ell_{(r)} (1-p)^{\ell-r} 
\label{derivs}
\eeq
for $1 \le r \le \ell$ and zero for $r \ge \ell+1$, where $\ell_{(r)}$ is the
falling factorial, 
\beq
\ell_{(r)} \equiv \prod_{s=0}^{r-1} (\ell-s) \ . 
\label{fallingfactorial}
\eeq

We will need to invert Eq. (\ref{pell}) and solve for $p$ in terms of
$p_\ell$.  For this purpose, we note that Eq. (\ref{pell}) is an $\ell$'th
degree algebraic equation for $p$, but the relevant one among the $\ell$ roots
is determined by the requirement that $v=0$ if and only if $v_\ell=0$, so $p=0$
if and only if $p_\ell=0$.  This root is given by
\beq
p = 1 - (1-p_\ell)^{1/\ell} \ . 
\label{pfrompell}
\eeq
Note that this relation has the same form as Eq. (\ref{pell}) with the
replacements $p \ \leftrightarrow \ p_\ell$ and $\ell \ \leftrightarrow \
1/\ell$.  

From Eqs. (\ref{pell}) and (\ref{pfrompell}) we derive the following
properties.  First, the transformation (\ref{pell}) maps the interval $p \in
[0,1]$ to the interval $p_\ell \in [0,1]$, and similarly the inverse
transformation (\ref{pfrompell}) maps the interval $p_\ell \in [0,1]$ to $p \in
[0,1]$. Second,
\beq
p = 0  \ \Longleftrightarrow \ p_\ell = 0
\label{ppp0}
\eeq
and 
\beq
p = 1  \ \Longleftrightarrow \ p_\ell = 1 \ . 
\label{ppp1}
\eeq
Third, for $\ell \ge 2$, 
\beq
p_\ell - p = (1-p)\, p_{\ell-1}
\label{pellminusp}
\eeq
Fourth, as is evident from Eq. (\ref{pellminusp}), 
\beq
p_\ell \ge p  \quad {\rm for} \ \ p \in [0,1] \quad {\rm and} \ \  \ell \ge 2 
\ , 
\label{pellpineq}
\eeq
with equality only at $p=p_\ell=0$ and $p=p_\ell=1$.  (For $\ell=1$, $p_1=p$,
so this inequality is realized as an equality for all $p$.)  For $\ell \ge 2$,
the difference $p_\ell - p$ has a maximum in the interval $0 \le p \le 1$ which
starts at $p=1/2$ for $\ell=2$ and moves to the left as $\ell$
increases.  Fifth, for fixed $p \in (0,1)$,
\beq
p_\ell \ {\rm is \ a \ monotonically \ increasing \ function \ of} \ \ \ell \ \
{\rm for} \ p \in (0,1) 
\label{pellmonotonicell}
\eeq
and for fixed $p_\ell \in (0,1)$, 
\beq
p \ {\rm is \ a \ monotonically \ decreasing \ function \ of} \ \ \ell \ \
{\rm for} \ p_\ell \in (0,1) \ .
\label{pmonotonicell}
\eeq
Sixth, from the $r=1$ special case of (\ref{derivs}), $dp_\ell/dp = \ell
(1-p)^{\ell-1}$, it follows that for fixed $\ell$, 
\beq
p_\ell \ {\rm is \ a \ monotonically \ increasing \ function \ of} \ \ p \ \
{\rm for} \ p \in (0,1) \ .
\label{pellmonotonicp}
\eeq
Similarly, for fixed $\ell$, 
\beq
p \ {\rm is \ a \ monotonically \ increasing \ function \ of} \ \ p_\ell \ \
{\rm for} \ p \in (0,1) \ .
\label{pmonotonicpell}
\eeq
For $\ell \ge 3$, the curve of $p_\ell$ as a function of $p$ is quite flat
near $p=1$ because, as is evident from Eq. (\ref{derivs}), the second
derivative $d^2p_\ell/dp^2$ vanishes at $p=1$. More generally, the $r$'th
derivative, $d^r p_\ell/dp^r$, vanishes at $p=1$ for $\ell \ge r+1$.

From Eqs. (\ref{zellhrel}) and (\ref{pell}), it follows that an
arbitrary percolation quantity $Q(p)_{\{ G \}_\ell}$ on $\{ G \}_\ell$, such 
as $\langle n \rangle_{\{ G \}_\ell}$, $P(p)_{\{ G \}_\ell}$, 
$S(p)_{\{ G \}_\ell}$, etc. satisfies the relation
\beq
Q(p)_{\{ G \}_\ell} = Q(p_\ell)_{\{ G \} } \ . 
\label{qprel}
\eeq
In particular, this relation holds for $N \to \infty$ limits of lattice graphs 
$\{ G \} = \Lambda$.  Thus, for
example, 
\beq
\langle n \rangle(p)_{\Lambda_\ell} = \langle n \rangle(p_\ell)_{\Lambda}  \ , 
\label{kell}
\eeq
\beq
P(p)_{\Lambda_\ell} = P(p_\ell)_{\Lambda} \ , 
\label{Pell}
\eeq
\beq
S(p)_{\Lambda_\ell} = S(p_\ell)_{\Lambda} \ , 
\label{Sell}
\eeq
\beq
\xi(p)_{\Lambda_\ell} = \xi(p_\ell)_{\Lambda} \ , 
\label{xiell}
\eeq
and so forth for other bond percolation quantities on the bond-inflated lattice
$\Lambda_\ell$, where $p_\ell$ is given in terms of $p$ by
Eq. (\ref{pell}). From either (\ref{Pell}) or (\ref{Sell}) in conjunction with
(\ref{pell}), it follows that 
\beq
p_{c,\Lambda_{\ell}} = 1-(1-p_{c,\Lambda})^{1/\ell}
\label{pclambdaell}
\eeq
and hence, in particular, 
\beq
p_{c,\Lambda_{\ell}} \le p_{c,\Lambda} \ , 
\label{pcinequality}
\eeq
with equality only for the case where $p_{c,\Lambda}=p_{c,\Lambda_\ell}=1$.
These relations (\ref{pell}), (\ref{pellmonotonicell}), (\ref{qprel}), and
(\ref{pclambdaell}) are important exact results, since they describe the effect
of the $\ell$-fold bond inflation on percolation quantities. This $\ell$-fold
bond inflation increases the vertex degree by the factor $\ell$ and has the
consequence that percolation quantities on $\{ G\}_\ell$ are equal to the
corresponding quantities on $\{ G \}$ with the bond occupation probability $p$
replaced by the larger probabiity $p_\ell$.  This means that as $p$ increases,
the infinite percolation cluster appears at a smaller value of $p$ on $\{
G\}_\ell$ than on $\{ G \}$, as given by Eq. (\ref{pclambdaell}).  That is, the
$\ell$-fold bond inflation enhances the formation of an infinite percolating
cluster on the resultant graph $\{ G\}_\ell$.  The relation (\ref{pclambdaell})
is specific to this process of $\ell$-fold bond inflation, while the general
inequality (\ref{pcinequality}) follows from the fact that if a lattice
$\Lambda'$ is obtained from a lattice $\Lambda$ by the addition of (an
arbitrary) set of bonds, then $p_{c,\Lambda'} \le p_{c,\Lambda}$.  Recall that
this fact is clear, since if $p$ is large enough for an infinite percolation
cluster to exist on $\Lambda$, then this cluster certainly also exists on
$\Lambda'$ with its additional bonds.

The critical percolation threshold depends not only on the lattice dimension
$d$ and vertex degree $\kappa$ for a regular $\kappa$-regular lattice $\Lambda$
or effective vertex degree $\kappa_{eff}$ for a lattice with vertices of
different degrees, but also on other related properties of the lattice. Several
studies of lattice properties that affect $p_{c,\Lambda}$ have been carried out
for regular lattices, including Archimedean lattices and their planar duals
\cite{essam80}-\cite{parv07}.  Clearly, another graphical property that is
relevant is the edge-connectivity (= bond-connectivity) $\lambda(G)$, defined
as the minimum number of bonds that must be removed to increase the number of
components by one.  In particular, given that $G$ is connected, i.e., $n(G)=1$,
the edge connectivity, $\lambda(G)$ is the minimum number of bonds that must be
removed to separate the graph into two disconnected components.  This is
related to, but different from, vertex degree measures such as $\kappa$ for a
$\kappa$-regular graph or $\kappa_{eff}$ for the limit $N \to \infty$ of a
graph with vertices of several different degrees.  For example, a tree graph
$G_t$ (defined as a connected graph with no circuits) may have vertices with
arbitrary degrees, but has $\lambda(G_t)=1$.  For a $\kappa$-regular lattice
with periodic boundary conditions, $\lambda(\Lambda)=\kappa$.  For our present
discussion, we note that
\beq
\lambda(G_\ell) = \ell \, \lambda(G) \ . 
\label{lambdagell}
\eeq
Thus, for an arbitrary graph $G$, the $\ell$-fold bond inflation increases both
the vertex degree and the edge-connectivity by the factor of $\ell$.

Although only the behavior of $p$ and $p_\ell$ in the interval $p \in [0,1]$
and thus $p_\ell \in [0,1]$ is of direct interest for percolation, the behavior
for real $p$ and $p_\ell$ outside this interval is also of interest in a
broader context.  We note that for $p < 0$, $p_\ell$ is also negative and for
$p >> 1$, $p_\ell$ is positive if $\ell$ is odd and negative if $\ell$ is
even. Furthermore, for $\ell \ge 2$, the point $p=1$ is a (global) maximum for
$p_\ell$ if $\ell$ is even and an inflection point if $\ell$ is odd.  Moreover,
in addition to its zero at $p=0$, $p_\ell$ vanishes at $p=2$ if and only if
$\ell$ is even.  Even more generally, as will be discussed below, the analytic
behavior of various percolation quantities in the complex $p$ plane is of
interest.  Indeed, there are cases where the radius of convergence of a Taylor
series expansion for $\langle n \rangle$ about the point $p=0$ is not set by
$p_c$, but instead by complex singularities in the $p$ plane. Explicit examples
of this were given in Ref. \cite{pc}.


\subsection{Percolation Threshold on Some Specific Bond-Inflated Lattices} 

We now focus on regular lattice graphs $\{G \} = \Lambda$ and $\{G_\ell \} =
\Lambda_\ell$.  It is of interest to apply our general result
(\ref{pclambdaell}) to obtain some illustrative numerical values of
$p_{c,\Lambda_\ell}$ on various lattices.  First, on a one-dimensional or
quasi-one-dimensional (Q1D) infinite-length, finite-width strip
$\Lambda_{Q1D}$, with $p_{c,Q1D}=1$, the $\ell$-fold bond inflation leaves this
property unchanged, i.e., $p_{c,\Lambda_{Q1D_\ell}}=1$, as is clear from
(\ref{pell}) and (\ref{ppp1}).  On higher-dimensional lattices, as special
cases of our general relation (\ref{pclambdaell}), we display the following
illustrative results: 
\beq
p_{c,(sq)_\ell} = 1 - (1-p_{c,sq})^{1/\ell} = 
  1 - \Big ( \frac{1}{2} \Big )^{1/\ell}
\label{pc_sq_ell}
\eeq
\beq
p_{c,(tri)_\ell} = 1 - (1-p_{c,tri})^{1/\ell} = 
   1 - \Big [ 1-2 \sin \Big ( \frac{\pi}{18} \Big ) \Big ]^{1/\ell}
\label{pc_tri_ell}
\eeq
\beq
p_{c,(hc)_\ell} = 1 - (1-p_{c,hc})^{1/\ell} = 
   1 - \Big [ 2 \sin \Big ( \frac{\pi}{18} \Big ) \Big ]^{1/\ell}
\label{pc_hc_ell}
\eeq
\beq
p_{c,(sc)_\ell} = 1 - (1-p_{c,sc})^{1/\ell} = 
   1 - (0.751187)^{1/\ell} \ . 
\label{pc_scc_ell}
\eeq
We list numerical values of these threshold percolation probabilities in Table
\ref{pcvalues}. Note that $\kappa((hc)_\ell)=3\ell$, $\kappa((sq)_\ell)=4\ell$,
and $\kappa((tri)_\ell)= \kappa((sc)_\ell) = 6\ell$. 
\begin{table}
\caption{\footnotesize{Values of the critical bond percolation threshold
probability $p_{c,\Lambda_\ell}$ on the lattice $\Lambda_\ell$ obtained from
the lattice $\Lambda$ by replacing each bond by $\ell$ bonds (so that
$\Lambda_1 \equiv \Lambda$). We list results for the square (sq), triangular
(tri), honeycomb (hc), and simple cubic (sc) lattices. See text for further
details.}}
\begin{center}
\begin{tabular}{|c|c|c|c|c|} \hline\hline
$\ell$ & $p_{c,(sc)_\ell}$ & $p_{c,(tri)_\ell}$ & $p_{c,(sq)_\ell}$
      & $p_{c,(hc)_\ell}$ \\
\hline
 1  &  0.249   &  0.347   &  0.500   &  0.653   \\
 2  &  0.133   &  0.192   &  0.293   &  0.411   \\
 3  &  0.0910  &  0.133   &  0.206   &  0.297   \\
 4  &  0.0690  &  0.101   &  0.159   &  0.232   \\
 5  &  0.0556  &  0.0818  &  0.129   &  0.191   \\
 6  &  0.0466  &  0.0686  &  0.109   &  0.162   \\
 7  &  0.0400  &  0.0591  &  0.0943  &  0.140   \\
 8  &  0.0351  &  0.0519  &  0.0830  &  0.124   \\
\hline\hline
\end{tabular}
\end{center}
\label{pcvalues}
\end{table}
%


\subsection{Effect of Bond Inflation on Universality Class of Percolation
  Transition} 

Consider a regular $d$-dimensional lattice $\Lambda$ with $d \ge 2$, so that
$p_{c,\Lambda} \in (0,1)$.  As noted above, the percolation transition is in
the same universality class as the phase transition in the ferromagnetic Potts
model in the limit $q \to 1$. The universality class of a finite-temperature
ferromagnetic phase transition depends only on the symmetry group of the
(zero-field) Hamiltonian and the dimensionality of the lattice.  In particular,
this universality class is independent of the coordination number of the
lattice.  From these facts it follows that the process of bond inflation does
not change the universality class of the percolation transition on such a
lattice.  In contrast, special properties apply in the case of a
one-dimensional lattice and infinite-length, finite-width lattice strips, which
are quasi-one-dimensional.  For these there is no finite-temperature phase
transition in a spin model (with short-range spin-spin interactions), and,
correspondingly, the critical percolation threshold is $p_c=1$.  We will show
below that for one-dimensional and quasi-one-dimensional lattices, bond
inflation can change a critical exponent characterizing percolation.


\subsection{Effect of Bond Inflation on Percolation on a 1D Lattice}

We denote the 1D lattices of length $N$ vertices and free and cyclic boundary
conditions as $L_n$ and $C_n$, respectively and the corresponding $\ell$-fold
bond-inflated lattices as $(L_n)_\ell$ and $(C_n)_\ell$.  The $N \to \infty$
limits of these lattices are denoted $\{ L \}$, $\{ C \}$, $\{ L \}_\ell$, and
$\{ C \}_\ell$.  Since percolation quantities are independent of the boundary
conditions we will denote both of these limits simply as $1D$, and 
the corresponding $\ell$-fold bond-inflated limit as $(1D)_\ell$.  It will be
convenient to recall how the well-known results for percolation on a
one-dimensional lattice follow from the solution to the Potts model in the $q
\to 1$ limit.  Since this model is only critical at $T=0$, i.e., 
$v=\infty$, it follows via Eq. (\ref{vp}) that 
\beq
p_{c,1D}=1 \ , 
\label{pc1d}
\eeq
Evaluating Eq. (\ref{kdfdq}) with the reduced free energy $f_{1D}=\ln(q+v)$,
one has 
\beq
\langle n \rangle_{1D} = 1-p \ , 
\label{k1d}
\eeq
For any $p \in [0,1)$, in the limit $N
\to \infty$, there is zero probability that a vertex is in an infinite cluster
because there is no infinite cluster. Such a cluster only exists for
$p=p_{c,1D}=1$.  Hence, 
\beq
P(p)_{1D} = \cases{ 0 & if $p \in [0,1)$ \cr
               1 & if $p=1$ \cr} \ , 
\label{P1d}
\eeq
 This is analogous to the singularity in the magnetization for
a 1D spin model, which is identically zero for any finite temperature and jumps
to 1 at the critical temperature, $T=0$.  

The 1D Potts model correlation function has the form 
\beq
G(r) \propto \rho^r \ ,
\label{gr}
\eeq
where $\rho$ is the ratio of the next-to-maximal eigenvalue of the 
transfer matrix to the maximal eigenvalue,
\beq
\rho = \frac{\lambda_{submax}}{\lambda_{max}} \ . 
\label{rho}
\eeq
Here, $\rho_{1D} =v/(q+v)$. Setting $q=1$ and $v=v_p$ in Eq. (\ref{vp}) gives 
$\rho=p$, so 
\beq
G(r)_{1D}=p^r \ , 
\label{gr1d}
\eeq
With $G(r) \sim e^{-r/\xi}$ for $r \to \infty$ and $p \ne p_c$, 
one has
\beq
\xi_{1D} = -\frac{1}{\ln p} 
\label{xi1d}
\eeq
and hence 
\beq 
\xi_{1D} \sim \frac{1}{1-p} \quad {\rm as} \ \ p \nearrow 1 \ . 
\label{xi1dsing}
\eeq
Therefore, the corresponding critical exponent is 
\beq
\nu_{1D}=1 \ . 
\label{nu1d}
\eeq
Similarly, 
\beq
S(p)_{1D} = \frac{1+p}{1-p} \ , 
\label{S1d}
\eeq
so that $S(p)_{1D}$ diverges as $p \nearrow 1$ with the critical exponent 
\beq
\gamma_{1D}=1 \ . 
\label{gamma1d}
\eeq

Having reviewed these well-known results, we now analyze the effects of
$\ell$-fold bond inflation. From Eqs. (\ref{pell}) or (\ref{pclambdaell}), it
follows that
\beq
p_{c,(1D)_\ell} = 1 \ .
\label{pc1dell}
\eeq
As special cases of our general result (\ref{qprel}), we have 
\beq
\langle n \rangle_{(1D)_\ell} = (1-p)^\ell \ , 
\label{k1dell}
\eeq
\beq
P(p)_{(1D)_\ell} = P(p)_{(1D)} = \cases{ 0 & if $p \in [0,1)$ \cr
               1 & if $p=1$ \cr} \ , 
\label{P1dell}
\eeq
and
\beq
S(p)_{(1D)_\ell} = \frac{2-(1-p)^\ell}{(1-p)^\ell}  \ , 
\label{S1dell}
\eeq
As $p \nearrow p_{c,1D}=1$, this diverges like $S(p) \sim 2/(1-p)^\ell$.
From this we find that the critical exponent for bond percolation on the
$\ell$-fold bond-inflated lattice $(1D)_\ell$ is 
\beq
\gamma_{_{(1D)\ell}} = \ell  \ .
\label{gamma1dell}
\eeq
Furthermore, 
\beq
G(r)_{(1D)_\ell}=(p_\ell)^r \ , 
\label{gr1dell}
\eeq
so 
\beq
\xi_{(1D)_\ell} = -\frac{1}{\ln p_\ell} 
\label{xi1dell}
\eeq
Therefore, 
\beq 
\xi_{(1D)_\ell} \sim \frac{1}{1-p_\ell} = \frac{1}{(1-p)^\ell} \ , 
\quad {\rm as} \ \ p \nearrow 1
\label{xi1dellsing}
\eeq
and hence the correlation length diverges as $p \nearrow 1$ on the $\ell$-fold
bond-inflated lattice $(1D)_\ell$ with critical exponent 
\beq
\nu_{(1D)_\ell}=\ell \ . 
\label{nu1dell}
\eeq
These are important results, because they show that, in contrast to
percolation on higher-dimensional lattices, where $p_c \in (0,1)$, here the
bond-inflation changes the critical exponents and hence the universality 
class of the percolation transition.

\subsection{Effect of Bond Inflation on a Quasi-One-Dimensional Lattice 
Strips}

Our result that $\ell$-fold bond inflation changes the universality class of
the percolation transition on a one-dimensional lattice generalizes to apply
also to quasi-one-dimensional, infinite-length, finite-width strips $\{ G_s
\}$. This is a consequence of the fact that, because the Potts ferromagnet is
only critical at $T=0$ (i.e., $v=\infty$) on such strips, corresponding to
$p_{c,\{ G_s \}}=1$, divergences of the form
\beq
S(p)_{\{ G_s \}} \propto \frac{1}{(1-p)^{\gamma_{\{ G_s \}}}}
\label{spgs}
\eeq
and
\beq
\xi(p)_{\{ G_s \}} \propto \frac{1}{(1-p)^{\nu_{\{ G_s \}}}}
\label{xigs}
\eeq
as $p \nearrow 1$ change to
\beq
S(p)_{\{G_s \}_\ell} \propto \frac{1}{(1-p_\ell)^{\gamma_{\{ G_s \}}}}
                      = \frac{1}{(1-p)^{\ell \gamma_{\{ G_s \}}}}
\label{spgsell}
\eeq
and
\beq
\xi(p)_{\{G_s \}_\ell } \propto \frac{1}{(1-p_\ell)^{\ell \gamma_{\{ G_s\}}}}
                        = \frac{1}{(1-p)^{\ell \nu_{\{ G_s \}}}} \ . 
\label{xigsell}
\eeq
Hence, the corresponding critical exponents on the $\ell$-fold bond-inflated
infinite-length, finite-width strip graph $\{ G_s \}_\ell$ are 
\beq
\gamma_{ \{ G_s \}_\ell} = \ell \gamma_{ \{ G_s \} } 
\label{gammagsell}
\eeq
and
\beq
\nu_{ \{ G_s \}_\ell} = \ell \nu_{ \{ G_s \} }  \ . 
\label{nugsell}
\eeq
Thus, again the universality class is changed by the $\ell$-fold bond
inflation on these infinite-length, finite-width strip graphs. 


\section{Percolation on Specific Infinite-Length Lattice Strip Graphs}

In this section we give some results for the infinite-length limits of lattice
strip graphs.  We begin with the $L_x \to \infty$ limit of the square-lattice
strip graph with width $L_y=2$ and free ($F$) transverse boundary conditions,
i.e., the ladder graph. The longitudinal boundary conditions are free or
periodic.  We denote this limit as $sq,2_F$, where the subscript $F$ refers to
the transverse boundary conditions. We have previously calculated the average
per-site cluster number, which is \cite{pc}
\beq
\langle n \rangle_{sq,2_F} = \frac{(1-p)^2(2+p-2p^2)}{2(1-p^2+p^3)} \ .
\label{nlad}
\eeq
A notable property of this exact result is that it has a pole singularity 
at a negative real value of $p$, namely $p=-0.7549$ (as well as two complex 
values of $p$) and, as is evident, the singularity at this negative real value
is closer to the origin than the critical percolation threshold value,
$p_c=1$. Hence, the radius of convergence of a small-$p$ Taylor series
expansion of this quantity is set not by $p_c$, but by this unphysical
singularity. This phenomenon of an unphysical singularity being closer to the
origin than the physical singularity is also true of our result in 
Eq. (\ref{kk2jn}) below and of many infinite-length, finite-width strips
analyzed in \cite{pc}.  

Here we go on to calculate the divergences in $\xi(p)$ and $S(p)$ for this
strip as $p \to p_c=1$.  Using the solution in Ref. \cite{a} for the general
Potts model partition function on a strip of this type of arbitrary length, we
calculate
\beq
\rho_{sq,2_F} = \frac{\lambda_{sq,2_F,submax}}{\lambda_{sq,2_F,max}} \ , 
\label{rholad}
\eeq
where
\begin{widetext}
\beq
\lambda_{sq,2_F,max} = \frac{1}{2} \bigg [ v^3+4v^2+3qv+q^2
+ \Big [ v^6+4v^5-2qv^4-2q^2v^3+12v^4+16qv^3+13q^2v^2+6q^3v+q^4 \Big ]^{1/2}
\bigg ] \label{lamladmax}
\eeq
and
\beq
\lambda_{sq,2_F,submax} = \frac{v}{2}\bigg [ q+v(v+4) +
\Big [ v^4+4v^3+12v^2-2qv^2+4qv+q^2 \Big ]^{1/2} \bigg ] \ . 
\label{lamladsubmax}
\eeq
\end{widetext}

In the limit $K \to \infty$, i.e., $a \to \infty$, this leads to a divergence
in $\xi$ of the form
\beq
\xi_{sq,2_F} \sim \frac{a^2}{q} \quad {\rm as} \ \ a \to \infty
\label{xilad}
\eeq
where $a=e^K$, as defined in (\ref{kdef}).  Setting $q=1$
and $v=v_p$ as in Eq. (\ref{vp}), we thus obtain
\beq
\xi_{sq,2_F} \sim \frac{1}{(1-p)^2} \quad {\rm as} \ \ p \nearrow 1
\label{xiladperc}
\eeq
Hence, the corresponding correlation-length critical exponent for bond 
percolation on the infinite ladder graph is 
\beq
\nu_{sq,2_F} = 2 \ . 
\label{nulad}
\eeq

By similar methods we derive the following results for some other
infinite-length limits of finite-width lattice strips. As before, our 
results apply for either free or periodic longitudinal boundary conditions.
First, we consider the strip of the square lattice with (transverse) width 
$L_y=2$, with periodic ($P$), rather than free, transverse boundary
conditions.  In effect, this doubles each transverse bond, leaving the
longitudinal bonds unchanged.    We denote the $L_x \to \infty$ limit of this
strip as $sq,2_P$.  $\langle n \rangle_{sq,2_P}$ was calculated in \cite{pc}. 
Using our results from \cite{s3a}, 
we calculate the divergence in the correlation length for the ferromagnetic
Potts model as $T \to 0$ on this strip to be 
\beq
\xi_{sq,2_P} \sim \frac{a^2}{q} \quad {\rm as } \ \ a \to \infty \ . 
\label{xitor}
\eeq
Setting $q=1$ and $v=v_p$, we derive the
corresponding results for the divergence in $\xi$ for bond percolation on
this strip:
\beq
\xi_{sq,2_P} \propto \frac{1}{(1-p)^2} \quad {\rm as} \ \ p \nearrow 1 \ , 
\label{xisqtor}
\eeq
which is the same as for the $sq,2_F$ strip. 
Hence, the correlation-length critical exponent for bond 
percolation on this strip is 
\beq
\nu_{sq,2_P} = 2 \ . 
\label{nusq2p}
\eeq

Next, we consider the width $L_y=2$ strip of the triangular lattice with free
transverse boundary conditions, denoted $tri,2_F$.  We calculated $\langle n
\rangle_{tri,2_F}$ in \cite{pc}.  Here, using our results in \cite{ta}, we
calculate the divergence in the correlation length for the Potts ferromagnet on
this strip as $T \to 0$ to be
\beq
\xi_{tri,2_F} \sim \frac{a^3}{2q} \quad {\rm as } \ \ a \to \infty
\label{xitri}
\eeq
and hence for the bond percolation problem,  
\beq
\xi_{tri,2_F} \propto \frac{1}{(1-p)^3} \quad {\rm as} \ \ p \nearrow 1 \ .
\label{xiladtri}
\eeq
Consequently, 
\beq
\nu_{tri,2_F} = 3 \ . 
\label{nutri}
\eeq

These results again show how the critical behavior of percolation is sensitive
to details of the lattice structure at the lower critical dimensionality.  Note
that $\kappa_{1D}=2$, $\kappa_{sq,2_F}=3$, $\kappa_{sq,2_P}=4$, and
$\kappa_{tri,2_F}=4$ for these strips with periodic longitudinal boundary
conditions. (More generally, for the corresponding strips with free
longitudinal boundary conditions, these values apply for the respective
$\kappa_{eff}$.)  Carrying out $\ell$-fold bond inflation on these
infinite-length strips, one again gets a change in the critical exponents and
hence universality class for the percolation transition, as a special case of
(\ref{nugsell}).


\section{Bond Percolation on $G[K_r,jn]$} 

\subsection{General} 

In this section we present exact expressions for the average cluster number
$\langle n \rangle$ for the infinite-length limits of a family of graphs with
variable vertex degree, namely $\lim_{m \to \infty} G[(K_r)_m,jn,BC]$, defined
as the infinite-length limit of a line or ring of $m$ subgraphs $K_r$ connected
in such a manner that all vertices of the $\ell$'th $K_r$ are connected to all
vertices of the $(\ell+1)$'th $K_r$. Here the notation BC refers to the
longitudinal boundary conditions, which are free (FBC) for the line and
periodic (PBC) for the ring.  We also present results for the divergence in the
correlation length.  One of the reasons that this family of graphs is
useful for the present study is that one can vary the vertex degree over a
rather wide range by varying $r$. An illustrative example of a member of the
family $G[(K_r)_m,jn,PBC]$ for the case $r=2$ and $m=4$ was given as Fig. 1(a)
in Ref. \cite{k}. The cyclic strip $G[(K_r)_m,jn,PBC]$ is a $\kappa$-regular
graph with uniform vertex degree
\beq
\kappa = 3r-1 \quad {\rm for} \ \ G[(K_r)_m,jn,PBC] \ , 
\label{kappakrjn}
\eeq
and all of the vertices except the end vertices of the free strip
$G[(K_r)_m,jn,FBC]$ also have this degree.  Thus $G[(K_r)_m,jn,PBC]$ is a
$(3r-1)$-connected graph.  For both FBC and PBC, the graph $G[(K_r)_m,jn,BC]$
has $N=mr$ vertices so that the infinite-length limit can be written
equivalently as $m \to \infty$ or $N \to \infty$. Our result for $\langle n
\rangle$ depends on $r$, but not on these boundary conditions.  We denote the
formal limit $m \to \infty$ of this family as
\beq
G[K_r,jn] \equiv \lim_{m \to \infty} G[(K_r)_m,jn]
\label{gkrjn}
\eeq
where we suppress the BC in the notation, since $\langle n \rangle$ is
independent of the boundary conditions.  For FBC, as well as PBC, in the 
$m \to \infty$ limit, we have 
\beq
\kappa_{eff} = 3r-1 \quad {\rm for} \ \ G[K_r,jn] \ . 
\label{kappaeffkrjn}
\eeq

Our method to calculate $\langle n \rangle$ in \cite{pc} and here is to apply
eq. (\ref{kdfdq}) in conjunction with exact results that we have computed for
the free energy of the Potts model on infinite-length, finite-width strips of
various lattices.  We will give explicit expressions for the cases $r=1,2,3$
and relevant properties of the result for general $r$.  We will also display
Taylor series expansions of the resultant $\langle n \rangle$ for $p$ near 0
and for $p$ near 1, in the latter case, using the expansion variable $s \equiv
1-p$. Our current work is a continuation of our previous calculations of
$\langle n \rangle$ for various families of graphs in Ref. \cite{pc}.  Other
studies of $\langle n \rangle$ for bond percolation include \cite{ziff97}.


\subsection{Calculations for $G[K_r,jn]$} 

For the calculation of $\langle n \rangle$, one needs the reduced free energy
for the $m \to \infty$ limit of the strip graph $G[(K_r)_m,jn,BC]$.  This is
simplest for the case of free boundary conditions.  In \cite{ka3} we determined
the general structural form of the Potts model partition function
$Z(G[(K_r)_m,jn,BC],q,v)$ for this family.  As we discussed in \cite{pc}, to
calculate $\langle n \rangle$ on the $N \to \infty$ limit of a one-parameter
family of recursive graphs, such as lattice strip graphs, using the free energy
of the Potts model on these graphs, one needs the dominant term contributing to
the free energy in the ferromagnetic region.  For these classes of
one-parameter graphs $G_m$ of length $m$ subunits, the free energy has the form
of a sum of $m$'th powers of certain functions, multiplied by certain
coefficients of degree $d$, ranging from 0 to a maximal degree $d$ depending on
the transverse structure of the strip but not its length.  The dominant term in
the ferromagnetic region arises from the degree $d=0$ sector, and is the same
independent of the longitudinal boundary condition, in accord with the
requirement that the thermodynamic behavior should be independent of the
boundary conditions in the infinite-length limit.  For the present case, we
proved that $Z(G[(K_r)_m,jn,PBC],q,v)$ has the general structural form
\cite{ka3}
\beq
Z(G[(K_r)_m,jn],q,v) = \sum_{d=0}^r \mu_d \sum_{i=1}^{n_T(r,d)}
(\lambda_{G[K_r,jn],d,i})^m \ , 
\label{zgsumclan}
\eeq
where 
\beq
n_T(r,d) = \sum_{j=1}^{r-d+1} {r-1 \choose j-1}
\label{ntrdgen}
\eeq
and the coefficient $\mu_d$ is a polynomial in $q$ of degree $d$ given by 
\beq
\mu_0=1 
\label{mud0}
\eeq
and
\beq
\mu_d = {q \choose d} - {q \choose d-1} = \frac{q_{(d-1)}(q-2d+1)}{d!}
\ \ {\rm for} \ \ 1 \le d \le r \ , 
\label{mud}
\eeq
where $q_{(j)}$ is the falling factorial defined in (\ref{fallingfactorial}). 
(The symbols $n_T(r,d)$ in Eqs. (\ref{zgsumclan}) and (\ref{ntrdgen}) above 
and $n_{Zh}(L_y,G_D,d)$ in Eq. (\ref{zsdgtran2}) follow the notation used 
in our earlier papers and should not be confused with the notation $n(G')$ for
cluster numbers.) 

With FBC, only the $\lambda_{G[K_r,jn],0,i}$ contribute. For general $r$, we 
have
\beq
f(G[K_r,jn],q,v) = \frac{1}{r} \ln[\lambda_{G[K_r,jn],0,max}] \ , 
\label{fgen}
\eeq
where $\lambda_{G[K_r,jn],0,max}$ denotes the $\lambda_{G[K_r,jn],0,j}$ of
maximal magnitude for ferromagnetic $v=v_p$, i.e., real positive $v$
corresponding to $p \in [0,1]$.  Thus,
\beq
\langle n \rangle_{G[K_r,jn]} = \frac{\partial f(G[K_r,jn],q,v_p)}{\partial q} 
\bigg |_{q=1} \ . 
\label{kgen}
\eeq

If $r=1$, the graph $G[(K_1)_m,jn,BC]$ is the path graph $L_m$ on $m$ vertices
for free boundary conditions, and the circuit graph $C_m$ for periodic
BC, which has already been discussed above. 


\subsection{$G[K_2,jn]$}

Here we calculate the average cluster number, per vertex and the correlation
length for the $m \to \infty$ limit of the family $G[(K_2)_m,jn,BC]$, denoted
as $G[K_2,jn]$. A graph in this family can also be regarded as a square-lattice
ladder graph with next-nearest-neighbor bonds.  From our calculation of the
partition function for this graph in \cite{ka}, we have
\beq
f(G[K_2,jn],q,v) = \frac{1}{2} \ln[\lambda_{G[K_2,jn],+}) 
\label{fk2jn}
\eeq
where 
\beq
\lambda_{G[K_2,jn],\pm} = \frac{1}{2} \Big [ T_{K_2} \pm \sqrt{R_{K_2}} \ 
\Big ]
\label{lamk2jn}
\eeq
with
\beq
T_{K_2} = v^5 + 5v^4 + 10v^3 + 12v^2 + 5qv + q^2
\label{tk2jn}
\eeq
and
\begin{widetext}
\beqs
R_{K_2} & = & v^{10} + 10v^9 + 45v^8 + 116v^7 + 196v^6 + 224v^5 + 144v^4 
 - 6v^6q - 10v^5q + 40 v^4q + 104v^3q \cr\cr
 & - & 2v^5q^2 - 10 v^4q^2 - 4v^3q^2 + 41v^2q^2 + 10vq^3 + q^4  \ . 
\label{rk2jn}
\eeqs
\end{widetext}
Using Eq. (\ref{kdfdq}), we calculate
\beq
\langle n \rangle_{G[K_2,jn]} 
 = \frac{(1-p)^4(2+3p-4p^2-p^4+p^5)}{2(1-2p^2+6p^3-6p^4+2p^5)} \ . 
\label{kk2jn}
\eeq

For small $p$, $\langle n \rangle_{G[K_2,jn]}$ has the Taylor series expansion
\beq
\langle n \rangle_{G[K_2,jn]} 
 = 1 - \frac{5}{2}p + 2p^3 + \frac{7}{2}p^4 - p^5 +O(p^6) \ . 
\label{kk2jntaylor}
\eeq
For $p$ near to 1, Eq. (\ref{kk2jn}) has the Taylor series expansion, in terms
of the expansion variable
\beq
s \equiv 1-p \ ,
\label{s}
\eeq
\beq
\langle n \rangle_{G[K_2,jn]}
 = \frac{1}{2}s^4 + 2s^5 - 2s^7 + 4s^8 + O(s^9) \ . 
\label{kk2jnrtaylor}
\eeq

For the calculation of the divergence in the correlation length we find, for
the ferromagnetic Potts model on this strip,
\beq
\xi_{G[K_2,jn]} \sim \frac{a^4}{q} \quad {\rm as} \ \ a \to \infty \ . 
\label{xik2jn}
\eeq
Hence, setting $q=1$ and $v=v_p$, the divergence in the corresponding $\xi$ in
the bond percolation problem is 
\beq
\xi_{G[K_2,jn]} \sim \frac{1}{(1-p)^4} \quad {\rm as} \ \ p \nearrow 1 \ , 
\label{xik2jnperc}
\eeq
so that
\beq
\nu_{G[K_2,jn]} = 4 \ . 
\label{nuk2jn}
\eeq

For the various infinite-length, finite-width $\kappa$-regular lattice strips
for which we have carried out calculations, except for the $L_y=2$
square-lattice strip with toroidal boundary conditions, which involves double
bonds in the tranverse direction, we find that the correlation length in the
ferromagnetic Potts model diverges like $\xi \propto a^{\kappa-1}/q$ as $a \to
\infty$.  Thus, setting $q=1$ and $v=v_p$, this yields, for the corresponding
bond percolation problem on these strips, $\xi \propto 1/(1-p)^{\kappa-1}$ as
$p \nearrow 1$.  Further calculations with wider strips and other lattice types
are necessary to determine how general this formula is.  The fact that $\xi$
diverges in the same way for the $L_y=2$ square-lattice ladder strip and the
$L_y=2$ toroidal square-lattice strip indicates that for these strips, doubling
the bonds along a direction orthogonal to the longitudinal direction does not
change the critical behavior of the bond percolation.  This is understandable,
since it is only the longitudinal direction in which the $L_x \to \infty$ and
the infnite percolation cluster forms for $p=1$.  Our general result in
Eq. (\ref{xiell}) shows that if not only the transverse bonds, but also the
longitudinal bonds are doubled, this changes $\nu=2$ to $\nu=4$.

As in our earlier work \cite{pc}, although in an analysis of physical
percolation one is primarily interested in real $p \in [0,1]$, it is also
useful to investigate the analytic properties of $\langle n \rangle$ more
generally in the complex $p$ plane.  The reason for this is that singularities
in complex $p$ can have an important influence on series expansions in small or
large $p$ \cite{essam80,pc}.  Here we observe that $\langle n \rangle$ has
singularities, which are simple poles, at the zeros of the denominator of Eq.
(\ref{kk2jn}).  There are five such poles, which we list below to the indicated
accuracy:
\beq
 \{ -0.418530 \ ,  \ \ 0.300885 \pm 0.674465i \ , \ \ 
1.408380 \pm 0.454693i \} 
\label{kk2jnpoles}
\eeq
Of these, the first is the closest to $p=0$ and determines the radius of
convergence of the small-$p$ Taylor series for $\langle n \rangle$. 

\begin{figure}
\begin{center}
\includegraphics[height=6cm]{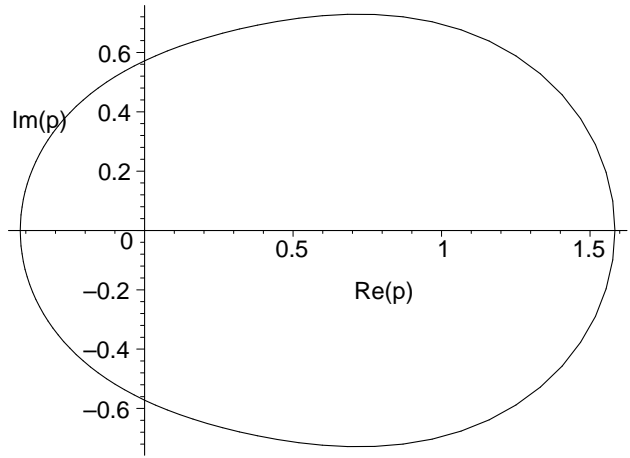}
\end{center}
\caption{\footnotesize{Plot of the boundary ${\cal B}_{qn}$ in the complex $p$
plane for the infinite-length limit $G[K_2,jn]$.}}
\label{K2plot}
\end{figure}

Thus, studies of percolation quantities on quasi-one-dimensional lattice strips
in \cite{pc} and here yield valuable insights into the influence of unphysical
singularities in series expansions about $p=0$ and $p=1$ for percolation on
higher-dimensional lattices \cite{comp}.  As in \cite{pc}, we can understand
these poles in $\langle n \rangle$ more deeply by noting that although the free
energy $f(\{G\},q,v)$ of a given infinite-length ($m \to \infty$) limit of a
lattice strip graph only depends on a dominant eigenvalue of the relevant
transfer matrix, the partition function for a finite strip graph $G_m$ is a sum
of $m$'th powers of these eigenvalues.  With the substitution $v=v_p$ in
Eq. (\ref{vp}), these eigenvalues become functions of $p$.  In the
infinite-length limit, partition function zeros in the complex $p$ plane merge
to form boundaries separating regions where a given eigenvalue is dominant,
i.e. has the largest magnitude and hence determines the resultant
$f(\{G\},q,v_p)$.  Plots of the resultant boundaries were shown for various
infinite-length limits of lattice graphs in \cite{pc}.  For the
$G[(K_2)_m,jn,FBC]$ strip, there are two such $\lambda$'s, given above in
Eq. (\ref{lamk2jn}).  Evaluating these for $v=v_p$ and $q=1$, we have
\beq
\lambda_{G[K_2,jn],+} = \frac{1}{(1-p)^5}
\label{lamk2jnplus}
\eeq
and
\beq
\lambda_{G[K_2,jn],-} = \frac{2p^2}{(1-p)^2} \ .
\label{lamk2jnminus}
\eeq
The boundary curve is the set of solutions of the
equation of degeneracy in magnitude of these $\lambda$s, namely 
\beq
2|p^2(1-p)^3|=1 \ .
\label{k2curve}
\eeq
The solution is a closed egg-shaped curve, shown in Fig. \ref{K2plot}.
This curve crosses the real $p$ axis at $p = -0.418530...$ (the only real root
of the equation $2r^2(1-r)^3-1=0$) and $p = 1.584080...$ (the only real root of
the equation $2r^2(1-r)^3+1=0$).  This curve thus constitutes the phase
boundary in the complex $p$ plane, separating this plane into two regions.  As
we proved in \cite{pc} for the infinite-length limit of an arbitrary strip
graph, the physical real interval $0 \le p \le 1$ lies entirely inside the
inner region bounded by this curve. The three poles of $\langle n \rangle$ in
Eq. (\ref{kk2jn}) lie on this boundary curve.


\subsection{$\langle n \rangle$ for $G[K_3,jn]$} 

One can also compute $\langle n \rangle$ for the $m \to \infty$ limit of the
family $G[(K_3)_m,jn,BC]$, viz., $G[K_3,jn]$.  From our calculation of the
partition function for this graph in \cite{ka3}, we obtain $f(G[K_3,jn],q,v) =
(1/3)\ln[\lambda_{G[K_3,jn],0,max}]$, where $\lambda_{G[K_3,jn],0,max}$ is the
root of maximal magnitude, in the relevant interval $p \in [0,1]$, of the cubic
equation displayed (for the Tutte polynomial equivalent to the Potts partition
function) in Eqs. (A.6)-(A.9) of \cite{ka3}.  The resulting expression for
$\langle n \rangle$ is too lengthy to include here, but we will give the
resultant Taylor series expansion for $p$ near 0,
\beq
\langle n \rangle_{G[K_3,jn]} = 
1 - 4p + \frac{19}{3}p^3 + 24p^4 + 39p^5 +O(p^6) \ . 
\label{kk3jntaylor}
\eeq

The small-$p$ Taylor series expansions of $\langle n \rangle_{ \{ G \} }$ have
the general form \cite{pc}
\beq
\langle n \rangle_{ \{ G \} } = 1 - \Big ( \frac{\kappa_{eff}}{2} \Big ) p + 
....
\label{ktaylorgeneral}
\eeq
where $\kappa_{eff}$ is the effective vertex degree for $\{ G \}$ and $...$
denote terms that are higher-order in $p$. Our Taylor series expansions of
$\langle n \rangle_{G[K_r,jn]}$ for the $r=2$ and $r=3$ cases, as well as the
elementary $r=1$ case are in accord with this general form, since $\kappa_{eff}
= 3r-1$ for these strips, as given in (\ref{kappaeffkrjn}). The increase in
$\kappa_{eff}$ with $r$ means that for a given $p$, there is an increased
probability of forming larger clusters, which, in turn, decreases the number of
clusters per vertex. 

These results on the infinite-length $G[K_r,jn]$ families containing $K_r$
subgraphs thus provide further insight into percolation on various families of
graphs.  Parenthetically, we note that rather than repeated $K_r$ subunits, one
could consider bond percolation on a single $K_N$ graph \cite{kr}.  The $K_N$
graph is much more highly connected than a graph in either of the families 
$G[(K_r)_m,jn,FBC]$ or $G[(K_r)_m,jn,PBC]$, since for $p=1$, each vertex of 
$K_N$ starts out connected to every other vertex.  Bond percolation on $K_N$
was studied in \cite{er} (reviewed in \cite{math}), and it was shown that (in a
probabilistic sense) the size of the largest connected component on $K_N$
diverges as $N \to \infty$ if $p \ge 1/N$.


\section{Analysis of Families of Graphs with Bounded Diameter} 

\subsection{Motivation} 

The essence of bond percolation on a usual lattice $\Lambda$ of dimension $d
\ge 2$ is that as the bond occupation probability increases through the
critical threshold value, $p_{c,\Lambda}$, an infinite percolation cluster
appears, linking vertices that are arbitrarily far apart on $\Lambda$.
Although $p_{c,\Lambda}=1$ for a one-dimensional or quasi-one-dimensional
lattice, the same statement applies.  One may ask how percolation quantities
would behave if one considered families of graphs that have the property of a
bounded diameter \cite{diam} as $N \to \infty$.  For these families of graphs,
the usual notion of a critical $p_c$ beyond which there is a percolation
cluster connecting two vertices arbitrarily far apart is clearly not
applicable. But how would the usual quanties such as $\langle n \rangle$,
$P(p)$, and $S(p)$ behave on such families of graphs?  Here we address this
question and calculate exact analytic expressions for these quantities on two
families of graphs with bounded diameter as $N \to \infty$. We note that, in
addition to the property of bounded diameter, both of these families consist of
planar graphs which also share a related property, namely that they both
contain a vertex whose degree goes to infinity as $N \to \infty$.  The second
family is also self-dual.

\subsection{Star Graphs} 

A star graph $S_N$ consists of one central vertex with degree $N-1$ connected
by bonds with $N-1$ outer vertices, each of which has degree 1.  (The context
will make clear the difference between this symbol and the symbol $S_N$ for the
symmetric group on $N$ objects.)  The graph $S_2$ is degenerate, in the sense
that it has no central vertex but instead coincides with $L_2$.  The graph
$S_3$ is nondegenerate, and coincides with $L_3$, while the $S_n$ for $N \ge 4$
are distinct graphs not coinciding with those of other families.  From the
calculation of $Z(S_n,q,v,w)$ in \cite{phs}, we have
\beq
f(\{ S \},q,v,w) = \ln(\lambda_{S})
\label{fstar}
\eeq
where
\beq
\lambda_{S} = q+w-1+wv \ .
\label{lamstar2}
\eeq
Setting $v=v_p$ and carrying out the differentiations to calculate $\langle
n \rangle$, $P(p)$, and $S(p)$, we obtain
\beq
\langle n \rangle_{\{S \}} = 1-p \ ,
\label{kstar}
\eeq
\beq
P(p)_{\{S \}} = p \ ,
\label{Pstar}
\eeq
and
\beq
S(p)_{\{S \}} = 1-p \ . 
\label{Sstar}
\eeq
We will comment on these results after calculating the corresponding quantities
for two other families of graphs. 

\subsection{Families of Self-Dual Graphs}

We next consider two families of planar self-dual (SD) graphs.  One family is
constructed by taking a path graph with $N-1$ vertices, adding one external
vertex, and connecting all of the vertices of the path graph to this external
vertex with single bonds, except for the vertex at one end, which is connected
to the external vertex with a double bond.  A second self-dual family is
constructed by taking a circuit graph $C_{N-1}$ with $N-1$ vertices, adding one
external vertex and connecting all of the $N-1$ vertices to this external
vertex, thereby forming the wheel graph $Wh_N$.  In \cite{sdg} we called these
dual boundary conditions (DBC) DBC1 and DBC2, and we calculated the Potts model
partition functions for them.  In the $N \to \infty$ limit, these yield the
same reduced free energy.  From this we computed \cite{pc}
\beq
\langle n \rangle_{(1D)_{SD}} = \frac{(1-p)^3}{1-p+p^2} \ .
\label{ksdg}
\eeq
(See Fig. 4 of \cite{pc} for a plot.)  

\begin{figure}
\begin{center}
\includegraphics[height=6cm]{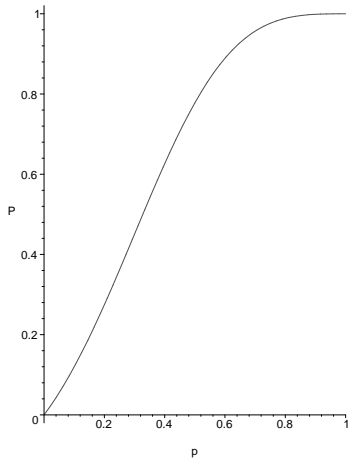}
\end{center}
\caption{$P(p)$ for the the $N \to \infty$ limit of the self-dual 1D graph, 
$(1D)_{SD}$. }
\label{Pwheelfig}
\end{figure}

\begin{figure}
\begin{center}
\includegraphics[height=6cm]{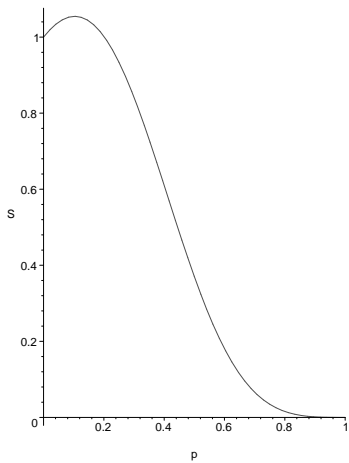}
\end{center}
\caption{$S(p)$ for the the $N \to \infty$ limit of the self-dual 1D graph, 
$(1D)_{SD}$. }
\label{Swheelfig}
\end{figure}

Here we generalize this analysis to the case of a finite external magnetic
field in order to calculate $P(p)$ and $S(p)$. As was true of the zero-field
case, in the relevant limit, namely $N \to \infty$, the free energy per vertex
is the same for DBC1 and DBC2 self-dual (SD) boundary conditions.  We give the
partition function for the case of self-dual boundary conditions of type 2
(DBC2), $Z(Wh_{n+1},q,v,w)$, in the appendix.  Taking the $N \to \infty$ limit
of this family, we calculate
\beq
f((1D)_{SD},q,v,w) = \ln[\lambda_{(1D)_{SD}}] \ , 
\label{fwheel}
\eeq
where $\lambda_{(1D)_{SD}} \equiv \bar\lambda_{Z,G_D,1,0,1}$ is given as the
solution with the $+$ sign in front of the square root in
Eq. (\ref{lamwheelquadratic}) of the appendix. Evaluating Eqs. (\ref{P}) and
(\ref{S}) for this case, we find
\beq
P(p)_{(1D)_{SD}} = \frac{p(1-p^2+p^3)}{(1-p+p^2)^2}
\label{Pwheel}
\eeq
and
\beq
S(p)_{(1D)_{SD}} = \frac{(1-p)^3(1+p-p^2)}{(1-p+p^2)^3} \ . 
\label{Swheel}
\eeq
These are plotted in Figs. \ref{Pwheelfig} and \ref{Swheelfig}. 

We now comment on the behavior of these quantities for both the $N \to \infty$
limit of the star graph family, $\{ S \}$ and of the self-dual graphs,
$(1D)_{SD}$. In accord with the general discussion given above, $\langle n
\rangle$ is a monotonically decreasing function of $p$, decreasing from
$\langle n \rangle=1$ at $p=0$ to $\langle n \rangle=0$ at $p=1$.  Recall that
$P(p)$ for the infinite 1D lattice vanishes identically for $p < p_c = 1$ and
has a jump discontinuity to the value $P(1)=1$. In contrast, $P(p)_{\{ S \}}$
and $P(p)_{(1D)_{SD}}$ are both nonzero and monotonically increasing in the
interval $p \in (0,1]$.  Moreover, for the 1D lattice, $S(p)$ diverges as $p$
approaches $p_c=1$ from below.  In contrast, $S(p)_{\{ S \} }$ decreases
monotonically from 1 to 0 as $p$ increases from 0 to 1, while for the same
range of $p$, $S(p)_{(1D)_{SD}}$, starts at 1, first increases, reaches a
maximum (of approximately 1.0545) at an intermediate point (namely,
$p=0.103657$, a root of the equation $dS(p)/dp=0$) and then decreases to zero
as $p \nearrow 1$.  The differences in behavior with respect to both the
infinite 1D lattice and higher-dimensional lattices show the effect of the 
fact that the $\{ S \}$ and $(1D)_{SD}$ families have bounded diameter. 
Among these families, one may also discern rather different analytic properties
in the complex $p$ plane.  The various quantities $\langle n \rangle$, $P(p)$
and $S(p)$ are entire functions for the $\{ S \}$ family.  In contrast, 
$\langle n \rangle$, $P(p)$, and $S(p)$ for the the infinite-length limit
$(1D)_{SD}$ have pole singularities at the same two complex-conjugate values of
$p$, namely
\beq
p=e^{\pm i \pi/3} \ , 
\label{psing}
\eeq
where $1-p+p^2$ vanishes.  The average per-site cluster number $\langle n
\rangle$ has single poles at each of these points, while $P(p)$ has double
poles, and $S(p)$ has triple poles.  In all cases, these singularities set the
radius of convergence of small-$p$ Taylor series about the origin as unity. It
is interesting that this is the same as the radius of convergence of the
small-$p$ series for $S(p)_{1D}$, but instead of a pole at $p=1$, one has here
a complex-conjugate pair of poles on the unit circle.


\section{Conclusions} 

In conclusion, in this paper we have presented some exact results on bond
percolation.  In one part of our work we have derived a relation between a
percolation quantity on an $\ell$-fold bond-inflated lattice $\Lambda_\ell$ and
the corresponding quantity on the original lattice $\Lambda$ evaluated with a
transformed bond occupation probability $p_\ell$ given by Eq. (\ref{pell}).
This is applicable for arbitrary lattices and, more generally, $N \to \infty$
limits of families of graphs, $\{ G \}$, and provides a precise measure of how
percolation quantities change as a consequence of this bond inflation.  As an
application of this general relation, we have calculated threshold bond
percolation probabilities on various bond-inflated lattices. We have shown that
this bond inflation leaves the universality class of the percolation transition
invariant for dimension $d \ge 2$ but changes it on a one-dimensional lattice
and on quasi-one-dimensional infinite-length, finite-width strips.  This was
demonstrated via changes in both the critical exponents $\gamma$ and $\nu$. We
have also presented expressions for the average cluster number $\langle n
\rangle$ per vertex for the bond percolation problem on the infinite-length
limits, $m \to \infty$, of a family of highly locally connected graphs, namely
$G[(K_r)_m,jn]$, for several $r$ values.  These add to one's knowledge of the
dependence of percolation quantities on vertex degree and also give further
insight into singularities of these quantities in the complex $p$ plane.
Finally, we have studied some families of graphs with bounded diameter and have
investigated, via analytic results, how this property affects quantities such
as $\langle n \rangle$, $P(p)$, and $S(p)$.

\begin{acknowledgments}

This research was partially supported by the Taiwan National Science Council
(NSC) grant NSC-100-2112-M-006-003-MY3 (S.-C. C.) and by the U.S. National
Science Foundation grant NSF-PHY-09-69739 (R.S.).

\end{acknowledgments}

\section{Appendix}

Here we give the partition function for the $q$-state Potts model on a 1D graph
with self-dual boundary conditions of type 2 (DBC2), which is a wheel graph
$Wh_n$ consisting of $n-1$ vertices forming a circuit graph $C_n$, each
connected by a bond to one central vertex. The general form of this partition
function for a DBC2 strip of length $L_x=m$ vertices and width $L_y$ vertices,
denoted $G_D,L_x \times L_y$ (and having $n=L_xL_y+1$ vertices in total) was
derived in (Eq. (7.17) of) Ref. \cite{zth}.  It is
\begin{widetext}
\beq
Z(G_D, L_y \times L_x,q,v,w) = \sum_{d=1}^{L_y+1} \tilde \kappa^{(d)}
\sum_{j=1}^{n_{Zh}(L_y,G_D,d)} (\lambda_{Z,G_D,L_y,d,j})^m + 
w \sum_{d=0}^{L_y} \tilde c^{(d)} \sum_{j=1}^{n_{Zh}(L_y,d)} (\bar
\lambda_{Z,G_D,L_y,d,j})^m \ , 
\label{zsdgtran2}
\eeq
\end{widetext}
where the numbers $n_{Zh}(L_y,G_D,d)$ and $n_{Zh}(L_y,d)$ were given for
general $L_y$ and $d$ in \cite{zth} and the coefficients 
$\tilde \kappa^{(d)}$ and $\tilde c^{(d)}$ are defined as follows: 
\beq
\tilde \kappa^{(d)} = \sum_{j=0}^{d-1} (-1)^j { 2d-1-j \choose j} (q-1)^{d-j} 
\label{kappatilde}
\eeq
and
\beq
\tilde c^{(d)} = \sum_{j=0}^d (-1)^j {2d-j \choose j}(q-1)^{d-j} \ . 
\label{cdtilde}
\eeq
In the general notation of \cite{zth}, the wheel graph is $Wh_N = G_D, L_x
\times L_y$ with $L_x=N-1$ and $L_y=1$. For the family of wheel graphs we thus
need the coefficients $\tilde \kappa^{(1)}=q-1$, $\tilde
\kappa^{(2)}=(q-1)(q-3)$, $\tilde c^{(0)}=1$, and $\tilde c^{(1)}=q-2$.  We
have (from Table 5 of \cite{zth}) $n_{Zh}(1,G_D,1)=3$, $n_{Zh}(1,G_D,2)=1$ and
(from Table 1 of \cite{zth}) $n_{Zh}(1,0)=2$, $n_{Zh}(1,1)=1$.  The three
$\lambda_{Z,G_D,1,1,j}$, $j=1,2,3$, are the roots of the following cubic
equation:
\beq
\xi^3 + a_2 \xi^2 + a_1 \xi + a_0 = 0 \ ,
\label{wheelcubic}
\eeq
where
\beq
a_2 = -v^2-3v-q+1-wv-w
\label{wheelcubic_a2}
\eeq
\beq
a_1 = v(qw+wv^2+4wv+q+w-1+v^2+qv)
\label{wheelcubic_a1}
\eeq
and
\beq
a_0 = -wv^2(v+1)(v+q) \ . 
\label{wheelcubic_a0}
\eeq
The two $\bar\lambda_{Z,G_D,1,0,j}$ are the roots of a quadratic equation,
\begin{widetext}
\beq
\bar\lambda_{Z,G_D,1,0,j} = \frac{1}{2} \bigg [ q+v-1+w(v+1)^2 
\pm \Big [ \{v+q-1+w(v+1)^2 \}^2 -4wv(v+1)(v+q) \Big ]^{1/2} \ \bigg ] \ . 
\label{lamwheelquadratic}
\eeq
where $j=1,2$ corresponds to the $\pm$ sign. 
Further, as a special case of the general structural results of \cite{zth}, 
\beq
\lambda_{Z,G_D,1,2,1} = v \ , \quad \bar\lambda_{Z,G_D,1,1,1}=v \ . 
\label{lamwheelv}
\eeq
Thus, explicitly, 
\beq
Z(Wh_N,q,v,w) = (q-1)\sum_{j=1}^3 (\lambda_{Z,G_D,1,1,j})^{N-1} 
+ (q-1)(q-3)v^{N-1} + w\sum_{j=1}^2  (\bar\lambda_{Z,G_D,1,0,j})^{N-1} + 
(q-2)wv^{N-1} \ . 
\label{zwheel}
\eeq
\end{widetext}
It is easily checked that in the special case of zero field, $w=1$, this
reproduces our calculation of $Z(Wh_N,q,v)$ in \cite{sdg}.  For the present
application, we only need the dominant $\lambda$ in the region $p \in [0,1]$,
i.e., via Eq. (\ref{vp}), $v \ge 0$, with $h \to 0^+$, i.e., $w \to 1^+$, and
this is $\bar\lambda_{Z,G_D,1,0,1}$. With the substitution $v=v_p$, $w \to
1^+$, we denote this simply as $\lambda_{(1D)_{SD}}$ in Eq. (\ref{fwheel}). 

Moreover, on the connection between percolation and the $q=1$ Potts model, we
note that from our exact calculation of the relevant transfer
matrix in \cite{zth} and the resultant $Z(L_n,q,v,w)$ and $Z(C_n,q,v,w)$ in
\cite{phs2}, it follows that 
\beq
f(\{L\},q,v,w) = f(\{C\},q,v,w) = \ln(\lambda_{L,1,0,+}) \ , 
\label{fline}
\eeq
where
\begin{widetext}
\beq
\lambda_{L,1,0,\pm} = \frac{1}{2} \Bigg [ q-1+v + w(1+v) \pm
 \Big [ \{q-1+v + w(1+v)\}^2-4vw(q+v)\Big ]^{1/2} \ \Bigg ] \ .
\label{lamcnplus}
\eeq
\end{widetext}
It is readily verified that applying the differentiations in (\ref{P}) and
(\ref{S}) reproduce the known results (\ref{P1d}) and (\ref{S1d}).  This
derivation is complementary to the usual one via combinatoric arguments.



\begin{thebibliography}{99}
\bibitem{loops}
%
We restrict our analysis here to graphs without any edges that connect a vertex
back to itself (called loops in mathematical graph theory) since these are 
not relevant to percolation.
\bibitem{kr}
%
The complete graph $K_r$ is defined as the graph with $r$ vertices such that
each vertex is connected by (single) bonds to each of the other vertices, so
that $K_r$ contains ${r \choose 2}$ bonds in all.
\bibitem{diam}
%
Consider a graph $G$ and any two vertices $i$ and $j$ on $G$.  Denote $d(i,j)$
as the distance (number of bonds) for a minimal-distance path connecting $i$
and $j$.  The diameter of $G$, denoted $d(G)$, is defined as the largest value
of $d(i,j)$ as $i$ and $j$ vary over $G$.
\end{thebibliography}
\end{document}